\documentclass[aps,pra,twocolumn,showpacs,superscriptaddress]{revtex4}
\usepackage{graphicx}
\usepackage{epstopdf}
\usepackage{amsmath}
\usepackage{amssymb}
\usepackage{float}
\usepackage{bm}

\input{epsf}
\begin{document}

\title{Hyperspherical asymptotics of a system of four charged particles}

\author{K.~M. Daily}
\affiliation{Department of Physics,
Purdue University,
West Lafayette, Indiana 47907, USA}

\date{\today}

\begin{abstract}
We present a detailed analysis of the charged four-body system in  
hyperspherical coordinates in the large hyperradial limit.
In powers of $R^{-1}$ for any masses and charges,
the adiabatic Hamiltonian is expanded to third order in the dimer-dimer limit 
and to first order in the particle-trimer limit.
\end{abstract}

\pacs{}

\maketitle

\section{Introduction}
The Born-Oppenheimer approach for diatomic molecules leads to
understandable one-dimensional potentials as a function of the
internuclear distance.
Generalizing to more degrees of freedom,
the hyperspherical coordinates defines a collective adiabatic coordinate,
the hyperradius $R$,
and treats all other degrees of freedom 
as ``fast'' coordinates~\cite{lin1995,nielsen2001}.
In this way,
we keep our intuitive Born-Oppenheimer-like picture
for ever larger systems,
though the challenge remains in solving the resulting adiabatic Hamiltonian
for each fixed value of $R$.

Many few-body systems have been analyzed using
the hyperspherical framework.
For example, 
nuclear systems~\cite{ballot1980,rosati1990}
utilizing an expansion in hyperspherical harmonics~\cite{avery1989,krivec1998},
systems of resonant short-range 
interactions~\cite{nielsen2001,vonstecher2009,rittenhouse2011,rakshit2012},
and systems of three 
charges~\cite{macek1968,lin1974,christensen1984,botero1986,macek1986,ho1987,
sadeghpour1990,archer1990,lin1995,esry2003,matveenko2009} 
to name several.
Hyperspherical methods have also been applied 
to molecular rearrangement collisions in physical chemistry, 
as discussed by Kuppermann for systems having up to five atoms 
in Ref.~\cite{kuppermann2011}, 
and references therein.
Charged systems in particular are interesting due to the long-range
nature of the interactionsand 
and because of their obvious relevance to many different physical processes 
in nature.
Also of note are several treatments of the four-body Coulomb problem 
with one massive nucleus and three electrons, 
specifically~\cite{watanabe1982,morishita1998,dincao2003}.  
And there has even been some work addressing atoms 
with more than three electrons, 
e.g. the five-body system consisting of a nucleus 
surrounded by four electrons treated in Ref.~\cite{morishita2005}.

The interplay between bound ionic complexes
in the field of the remaining charges 
is complicated to describe on an equal footing.
Moreover, 
the different types of fragmentation in molecules 
are usually treated entirely differently, 
with a separate class of methods for ionization channels 
of the H$_2$ molecule, 
for instance, 
than are utilized for molecular dissociation into neutrals or ions.  
This is one reason why a hyperspherical description could be fruitful, 
in that all fragmentation processes can then be mapped 
into a single coordinate $R$ approaching infinity.
The leading-order behavior in $R$ also shows the dominant long-range
potentials,
e.g. van der Waals or ionic,
that characterize such systems.

The present study generalizes the three-body asymptotic analysis
of Macek~\cite{macek1968},
considering here the four-body system of charges in the asymptotic limit.
We consider the dimer-dimer and particle-trimer asymptotic arrangements.
In hyperspherical coordinates,
the adiabatic Hamiltonian is expanded in powers of $R^{-1}$,
taking the dimer-dimer limit to third order
and the particle-trimer limit to first order.
Each term in the expansion is identified with analogous
terms of a multipole expansion between charged clusters.
The rest of the paper is organized as follows:
Sec.~\ref{sec_theory} defines the Hamiltonian
and the conversion to hyperspherical coordinates.
Section~\ref{sec_dd} focuses on the dimer-dimer configuration
while Sec.~\ref{sec_pt} focuses on the particle-trimer configuration.
Section~\ref{sec_conclusions} concludes.

\section{Theoretical background}
\label{sec_theory}
Consider the four-body system 
in three dimensions interacting via a sum of two-body Coulomb potentials.
The Hamiltonian $H$ in atomic units $(\hbar=m_e=1)$ reads
\begin{align}
H = & 
\sum_{j=1}^4 
\frac{-1}
     {2m_j} 
\nabla^2_{\bm{r}_j}
+\sum_{i<j}
\frac{q_i q_j}
     {|\bm{r}_i-\bm{r}_j|}
\end{align}
where $\bm{r}_j$, $m_j$, and $q_j$ 
are the location, mass, and charge of particle $j$.
The center of mass $H_{\rm CM}$ 
and relative $H_{\rm rel}$ contributions separate,
$H = H_{\rm CM} + H_{\rm rel}$
and our interest centers on the relative Hamiltonian,
\begin{align}
H_{\rm rel}  = 
  - \frac{1}{2 \mu} \sum_{j=1}^3 \nabla^2_{\bm{\rho}_j}
  + V_C(\bm{\rho}_1,\bm{\rho}_2,\bm{\rho}_3),
\end{align}
where $V_C$ contains the pair-wise Coulomb interactions as a function of
the three relative Jacobi vectors $\bm{\rho}_j$,
$j=1,2,3$.
The Jacobi vectors are scaled such that they are analogous to 
three equal-mass ``particles'' of mass $\mu$.
The hyperradial mass $\mu$ is arbitrary,
but for a unitary transformation
\begin{align}
\mu=\left(\frac{m_1m_2m_3m_4}{m_1+m_2+m_3+m_4}\right)^{1/3},
\end{align}
which is assumed for the rest of this work.
Under this transformation the volume element $dV$ 
associated with the relative degrees of freedom is
$dV= d\bm{\rho}_1d\bm{\rho}_2d\bm{\rho}_3$.

The relative Hamiltonian $H_{\rm rel}$ is recast in hyperspherical
coordinates in terms of eight hyperangles denoted by $\bm{\Omega}$
and a single length, the hyperradius $R$.
The relative Hamiltonian is then a sum of the 
hyperradial kinetic energy ${\cal T}_{R}$,
the hyperangular kinetic energy ${\cal T}_{\bm{\Omega}}$, and
the interaction potential,
\begin{align}
\label{eq_SErel}
H_{\rm rel} = {\cal T}_R + {\cal T}_{\bm{\Omega}} + V_{\rm int}(R,\bm{\Omega}).
\end{align}
Here,
\begin{align}
{\cal T}_R = - \frac{1}{2 \mu} 
\frac{1}{R^{8}}\frac{\partial}{\partial R}R^{8} 
\frac{\partial}{\partial R}
\end{align}
and
\begin{align}
{\cal T}_{\bm{\Omega}} = \frac{\bm{\Lambda}^2}{2\mu R^2},
\end{align}
where $\bm{\Lambda}$ is the grand angular momentum operator.
The exact form of the interaction potential $V_{\rm int}$
and the square of the grand angular momentum operator
depend on the choice of Jacobi vectors
and the hyperangles. 
In Secs.~\ref{sec_dd} and~\ref{sec_pt}
we choose the H-type and K-type trees, respectively~\cite{rittenhouse2011}.

The Jacobi vectors $\bm{\rho}_j$ are recast 
in terms of the hyperradius $R$
and the eight hyperangles $\alpha$, $\beta$, $\theta_1$, $\phi_1$,
$\theta_2$, $\phi_2$, $\theta_3$, and $\phi_3$.
Here,
$\theta_j$ and $\phi_j$ are defined as the usual polar and azimuthal angles of
the Jacobi vector $\bm{\rho}_j$
($0\le \theta_j \le \pi$ and $0\le \phi_j \le 2\pi$)
while
the hyperangles $\alpha$ and $\beta$ are defined via
\begin{align}
\label{eq_tanalpha}
\tan \alpha = \frac{\rho_1}{\rho_2}, 
\end{align}
and
\begin{align}
\tan \beta = \frac{\sqrt{\rho_1^2 + \rho_2^2}}{\rho_3}
\end{align}
where
\begin{align}
\rho_1 = & R \sin \alpha \sin \beta, \\
\rho_2 = & R \cos \alpha \sin \beta, \\
\label{eq_rho3}
\rho_3 = & R \cos \beta,
\end{align}
and $R^2 = \rho_1^2 + \rho_2^2 + \rho_3^2$.
The ranges of the hyperangles $\alpha$ and $\beta$ 
are restricted to $0\le \beta \le \pi /2$ 
and $0\le \alpha \le \pi /2$ because the $\rho_i$ are all positive.
With these definitions,
the square of the grand angular momentum operator is
\begin{align}
\label{eq_lambda2unscaled}
\bm{\Lambda}^2 = & 
- \bigg( \frac{\partial^2}{\partial \beta^2}
+ \left[\frac{5}{\tan \beta} - 2\tan \beta \right] 
     \frac{\partial}{\partial \beta}
\\ \nonumber
& \qquad 
+ \frac{1}{\sin^2 \beta} \left[ \frac{\partial^2}{\partial \alpha^2} 
  + \frac{4}{\tan (2 \alpha)} \frac{\partial}{\partial \alpha} \right]
  \bigg)
\\ \nonumber
& 
+ \frac{\bm{L}_1^2}{\sin^2 \alpha\sin^2 \beta}
+ \frac{\bm{L}_2^2}{\cos^2 \alpha\sin^2 \beta}
+ \frac{\bm{L}_3^2}{\cos^2 \beta},
\end{align}
where
\begin{align}
\bm{L}_j^2 = -\left(\frac{\partial^2}{\partial \theta_j^2}
+ \cot \theta_j \frac{\partial}{\partial \theta_j}
+ \frac{1}{\sin^2 \theta_j} \frac{\partial^2}{\partial \phi_j^2}\right)
\end{align}
is the square of the angular momentum operator with eigenvalue $l_j(l_j+1)$
associated with the Jacobi vector $\bm{\rho}_j$.
The volume element is
$dV= R^8 dR \,d\bm{\Omega}$,
where
$\,d\bm{\Omega}=\sin^2\alpha \cos^2\alpha \sin^5 \beta \cos^2\beta 
d\alpha d\beta d\hat{\bm{\rho}}_1d\hat{\bm{\rho}}_2d\hat{\bm{\rho}}_3 $
and $d\hat{\bm{\rho}}_j=\sin \theta_j d\theta_jd\phi_j$.

The solution $\Psi_E(R,\bm{\Omega})$ 
to Eq.~\eqref{eq_SErel} is expanded in terms of the scaled
radial $\bar{F}_{E\nu}(R)$ and 
scaled channel functions $\bar{\Phi}_{\nu}(R;\bm{\Omega})$,
\begin{align}
\label{eq_psi_expansion}
\Psi_E(R,\bm{\Omega}) = 
\sum_{\nu} F_{E\nu}(R) \Phi_{\nu}(R;\bm{\Omega}),
\end{align}
where
\begin{align}
F_{E\nu}(R) = R^{-4} \bar{F}_{E\nu}(R) 
\end{align}
and
\begin{align}
\label{eq_phiscaled}
\Phi_{\nu}(R;\bm{\Omega})=
\left(\sin \alpha \cos \alpha \sin^2 \beta \cos \beta\right)^{-1}
\bar{\Phi}_{\nu}(R;\bm{\Omega}).
\end{align} 
The channel functions at a fixed hyperradius $R$ form
a complete orthonormal set over the hyperangles,
\begin{align}
\label{eq_overlap}
  \int \,d\bar{\bm{\Omega}} \; 
     \bar{\Phi}^*_{\nu}(R;\bm{\Omega}) \bar{\Phi}_{\nu'}(R;\bm{\Omega}) 
  = \delta_{\nu \nu'},
\end{align}
and are the solutions to the adiabatic Hamiltonian $H_{\rm ad}(R,\bm{\Omega})$,
\begin{align}
  H_{\rm ad}(R,\bm{\Omega}) \bar{\Phi}_{\nu}(R;\bm{\Omega}) 
  = U_{\nu}(R) \bar{\Phi}_{\nu}(R;\bm{\Omega}),
\end{align}
where
\begin{align}
\label{eq_Had}
H_{\rm ad}  = & 
\frac{\bar{\bm{\Lambda}}^2}{2\mu R^2} 
+ \frac{C(\bm{\Omega})}{R}
\end{align}
and $C(\bm{\Omega})$ is the hyperangular part of the Coulomb interaction.
The scaled form of the square of the grand angular momentum operator 
$\bar{\bm{\Lambda}}^2$ in the form that acts directly on $\bar{\Phi}$ is
\begin{align}
\label{eq_lambda2}
\bar{\bm{\Lambda}}^2 = & -\left( \frac{\partial^2}{\partial \beta^2}
+ \frac{1}{\tan \beta} \frac{\partial}{\partial \beta}
+ \frac{1}{\sin^2 \beta} \frac{\partial^2}{\partial \alpha^2}\right)
\\ \nonumber
& 
+ \frac{\bm{L}_1^2}{\sin^2 \alpha\sin^2 \beta}
+ \frac{\bm{L}_2^2}{\cos^2 \alpha\sin^2 \beta}
+ \frac{\bm{L}_3^2}{\cos^2 \beta}.
\end{align}
Note that a term $4/\sin^2 \beta$ from scaling in $\alpha$ cancels
with a similar term from scaling in $\beta$.
Also a term of $12$ from scaling in $\beta$ cancels with 
a similar term from scaling in the hyperradius $R$.
The scaled volume element $d\bar{V}$ is
$\,d\bar{V}=\,dR \,d\bar{\bm{\Omega}}$ where
\begin{align}
d\bar{\bm{\Omega}}= 
\sin \beta 
d\alpha d\beta d\hat{\bm{\rho}}_1d\hat{\bm{\rho}}_2d\hat{\bm{\rho}}_3.
\end{align}

After applying the relative Hamiltonian Eq.~\eqref{eq_SErel} on the expansion 
Eq.~\eqref{eq_psi_expansion}
and projecting from the left onto the complete set of channel functions,
the Schr\"odinger equation reads
\begin{align}
\label{eq_SE_W}
&
\left( 
  -\frac{1}{2\mu}\frac{d^2}{d R^2} 
  + U_{\nu}(R)  -E
\right) F_{E \nu}(R) 
\\ \nonumber & \qquad 
-\frac{1}{2\mu} \sum_{\nu'}
  \left( 2 P_{\nu \nu'}(R)\frac{d}{d R} + Q_{\nu \nu'}(R) \right)
  F_{E \nu'}(R)
= 0.
\end{align}
The hyperspherical Schr\"odinger equation Eq.~\eqref{eq_SE_W}
is solved in a two step procedure.
First, $H_{\rm ad}(R,\bm{\Omega})$ is solved parametrically in $R$ for
the adiabatic potential curves $U_{\nu}(R)$.
In a second step,
the coupled set of one-dimensional equations in $R$ are solved. 
In Eq.~\eqref{eq_SE_W}, $P_{\nu\nu'}$ and $Q_{\nu\nu'}$ 
represent the coupling between channels,
where 
\begin{align}
\label{eq_Pcoupling}
  P_{\nu \nu'}(R) = \bigg\langle \bar{\Phi}_{\nu} \bigg| 
  \frac{\partial \bar{\Phi}_{\nu'}}{\partial R} \bigg\rangle_{\bar{\bm{\Omega}}}
\end{align}
and
\begin{align}
\label{eq_Qcoupling}
  Q_{\nu \nu'}(R) = \bigg\langle \bar{\Phi}_{\nu} \bigg| 
  \frac{\partial^2 \bar{\Phi}_{\nu'}}{\partial R^2} 
  \bigg\rangle_{\bar{\bm{\Omega}}}.
\end{align}
The brackets indicate that the integrals are taken only over the
hyperangles e.g. like in Eq.~\eqref{eq_overlap}, 
with the hyperradius $R$ held fixed.

\section{Dimer-dimer configuration}
\label{sec_dd}
\subsection{Coordinate system and Coulomb interaction}
\label{subsec_dd_cs}
To describe the fragmentation into two dimers 
at large hyperradius $R$,
it is convenient to choose the H-type set of Jacobi vectors.
The relative coordinates 
are defined via the coordinate transformation,
conveniently written in matrix form, as
\begin{align}
\label{eq_jmatrix}
\begin{pmatrix} 
\bm{\rho}_1 \\ \bm{\rho}_2 \\ \bm{\rho}_3 \\ \bm{R}_{CM}
\end{pmatrix} = 
\left( \!\!\!\!
\begin{array}{rlccr}
\sqrt{\frac{\mu_{12}}{\mu}}  \times & \{ 1 & -1 & 0 & 0 \} \\
\sqrt{\frac{\mu_{34}}{\mu}}  \times &\{ 0 & 0 & 1  & -1 \} \\
\sqrt{\frac{{\mu_{dd}}}{\mu}}\times &\{\frac{\mu_{12}}{m_2} 
                                   & \frac{\mu_{12}}{m_1} 
                                   & -\frac{\mu_{34}}{m_4} 
                                   & -\frac{\mu_{34}}{m_3} \} \\
\frac{1}{M_4} \times &\{ m_1 & m_2 & m_3 & m_4 \}
\end{array} \right) \!\!
\begin{pmatrix}
\bm{r}_1 \\ \bm{r}_2 \\ \bm{r}_3 \\ \bm{r}_4
\end{pmatrix} \!\! ,
\end{align}
where first $\bm{r}_1$ is joined to $\bm{r}_2$ (which defines $\bm{\rho}_1$) 
and $\bm{r}_3$ is joined to $\bm{r}_4$ (which defines $\bm{\rho}_2$),
then the center of mass of each subcluster is joined to define $\bm{\rho}_3$.
The reduced masses are 
$\mu_{12}=m_1m_2/(m_1+m_2)$,
$\mu_{34}=m_3m_4/(m_3+m_4)$, and
${\mu_{dd}}=(m_1+m_2)(m_3+m_4)/M_4$.
Also,
\begin{align}
\label{eq_massn}
M_N = m_1 + \cdots + m_N,
\end{align}
that is,
$M_N$ is the total mass of the first $N$ particles.
The following considers only dimer-dimer channels of the type (12)+(34), 
but the same analysis can be similarly developed 
for the other fragmentation possibilities.

The Coulomb part $V_C$ is straightforward to calculate
in Jacobi coordinates,
where the matrix from Eq.~\eqref{eq_jmatrix} is first inverted to define
the $\bm{r}_j$ in terms of the $\bm{\rho}_j$.
With the $\bm{r}_j$ defined, taking vector differences is straightforward.
\begin{widetext}
\begin{align}
\label{eq_Vc}
V_C & = 
   q_1q_2 \frac{\sqrt{\mu_{12} / \mu}}{\rho_1}
  +q_1q_3 \left|  \frac{\sqrt{\mu\;\mu_{12}}}{m_1}\bm{\rho}_1 
                - \frac{\sqrt{\mu\;\mu_{34}}}{m_3}\bm{\rho}_2 
                + \sqrt{\frac{\mu}{\mu_{dd}}}\bm{\rho}_3\right|^{-1}
  +q_1q_4 \left|  \frac{\sqrt{\mu\;\mu_{12}}}{m_1}\bm{\rho}_1 
                + \frac{\sqrt{\mu\;\mu_{34}}}{m_4}\bm{\rho}_2 
                + \sqrt{\frac{\mu}{\mu_{dd}}}\bm{\rho}_3\right|^{-1} 
\nonumber \\
& +q_2q_3 \left|  \frac{\sqrt{\mu\;\mu_{12}}}{m_2}\bm{\rho}_1 
                + \frac{\sqrt{\mu\;\mu_{34}}}{m_3}\bm{\rho}_2 
                - \sqrt{\frac{\mu}{\mu_{dd}}}\bm{\rho}_3\right|^{-1} 
  +q_2q_4 \left|  \frac{\sqrt{\mu\;\mu_{12}}}{m_2}\bm{\rho}_1 
                - \frac{\sqrt{\mu\;\mu_{34}}}{m_4}\bm{\rho}_2 
                - \sqrt{\frac{\mu}{\mu_{dd}}}\bm{\rho}_3\right|^{-1} 
  +q_3q_4 \frac{\sqrt{\mu_{34} / \mu}}{\rho_2}
\end{align}
Recasting the above expression in hyperspherical coordinates yields
\begin{align}
\label{eq_C}
C(\bm{\Omega}) = 
& q_1q_2 \frac{\sqrt{\mu_{12} / \mu}}{\sin \alpha\sin \beta} + \nonumber \\
&+q_1q_3 \bigg[    \frac{\mu\;\mu_{12}}{m_1^2}\sin^2 \alpha\sin^2 \beta 
                 + \frac{\mu\;\mu_{34}}{m_3^2}\cos^2 \alpha\sin^2 \beta 
                 + \frac{\mu}{{\mu_{dd}}}\cos^2 \beta
                 - \frac{\mu\sqrt{\mu_{12}\mu_{34}}}{m_1m_3} 
                        \sin(2\alpha)\sin^2 \beta\cos \theta_{12} \nonumber \\
& \qquad\qquad   + \frac{\sqrt{\mu_{12}}\;\mu}{\sqrt{{\mu_{dd}}}\;m_1}
                        \sin \alpha\sin(2\beta)\cos \theta_{13}
                 - \frac{\sqrt{\mu_{34}}\;\mu}{\sqrt{{\mu_{dd}}}\;m_3}
                        \cos \alpha\sin(2\beta)\cos \theta_{23}
         \bigg]^{-1/2}
\nonumber \\
&+q_1q_4 \bigg[    \frac{\mu\;\mu_{12}}{m_1^2}\sin^2 \alpha\sin^2 \beta 
                 + \frac{\mu\;\mu_{34}}{m_4^2}\cos^2 \alpha\sin^2 \beta 
                 + \frac{\mu}{{\mu_{dd}}}\cos^2 \beta
                 + \frac{\mu\;\sqrt{\mu_{12}\mu_{34}}}{m_1m_4} 
                        \sin(2\alpha)\sin^2 \beta\cos \theta_{12} \nonumber \\
& \qquad\qquad   + \frac{\sqrt{\mu_{12}}\;\mu}{\sqrt{{\mu_{dd}}}\;m_1}
                        \sin \alpha\sin(2\beta)\cos \theta_{13}
                 + \frac{\sqrt{\mu_{34}}\;\mu}{\sqrt{{\mu_{dd}}}\;m_4}
                        \cos \alpha\sin(2\beta)\cos \theta_{23}
         \bigg]^{-1/2}
\nonumber \\
&+q_2q_3 \bigg[    \frac{\mu\;\mu_{12}}{m_2^2}\sin^2 \alpha\sin^2 \beta 
                 + \frac{\mu\;\mu_{34}}{m_3^2}\cos^2 \alpha\sin^2 \beta 
                 + \frac{\mu}{{\mu_{dd}}}\cos^2 \beta
                 + \frac{\mu\;\sqrt{\mu_{12}\mu_{34}}}{m_2m_3} 
                        \sin(2\alpha)\sin^2 \beta\cos \theta_{12} \nonumber \\
& \qquad\qquad   - \frac{\sqrt{\mu_{12}}\;\mu}{\sqrt{{\mu_{dd}}}\;m_2}
                        \sin \alpha\sin(2\beta)\cos \theta_{13}
                 - \frac{\sqrt{\mu_{34}}\;\mu}{\sqrt{{\mu_{dd}}}\;m_3}
                        \cos \alpha\sin(2\beta)\cos \theta_{23}
         \bigg]^{-1/2}
\nonumber \\
&+q_2q_4 \bigg[    \frac{\mu\;\mu_{12}}{m_2^2}\sin^2 \alpha\sin^2 \beta 
                 + \frac{\mu\;\mu_{34}}{m_4^2}\cos^2 \alpha\sin^2 \beta 
                 + \frac{\mu}{{\mu_{dd}}}\cos^2 \beta
                 - \frac{\mu\;\sqrt{\mu_{12}\mu_{34}}}{m_2m_4} 
                        \sin(2\alpha)\sin^2 \beta\cos \theta_{12} \nonumber \\
& \qquad\qquad   - \frac{\sqrt{\mu_{12}}\;\mu}{\sqrt{{\mu_{dd}}}\;m_2}
                        \sin \alpha\sin(2\beta)\cos \theta_{13}
                 + \frac{\sqrt{\mu_{34}}\;\mu}{\sqrt{{\mu_{dd}}}\;m_4}
                        \cos \alpha\sin(2\beta)\cos \theta_{23}
         \bigg]^{-1/2}
\nonumber \\
&+q_3q_4 \frac{\sqrt{\mu_{34} / \mu}}{\cos \alpha\sin \beta}
\end{align}
\end{widetext}
where $\cos\theta_{ij}$ is the cosine of the angle between Jacobi vectors
$\bm{\rho}_i$ and $\bm{\rho}_j$ and 
the identity $2\sin x\cos x=\sin(2x)$ is used to simplify the expression.

\subsection{Effective Hamiltonian}
\label{subsec_dd_eh}
The simultaneous limits $R\to\infty$ and $\beta \to 0$,
such that $R\beta = \rho$,
describe the situation where the centers of mass of 
the (1,2) and (3,4) dimers are far apart.
Said differently, to lowest order
the Jacobi vector $\rho_3$ that connects the centers
of mass of the (1,2) and (3,4) dimers scales proportionally 
to the hyperradius R.

The asymptotic expansion of $H_{\rm ad}$ is accomplished
by setting $\beta=\rho/R$ 
in Eqs.~\eqref{eq_lambda2} and~\eqref{eq_C}
and taking the asymptotic Taylor series 
around $R \to \infty$.
Inverse powers of the hyperradius $R$ then count the order of the expansion.
This expansion defines the effective Hamiltonian $H_{\rm eff}$
\begin{align}
\label{eq_expand2}
H_{\rm eff} \approx H_0 + H_1 + H_2 + H_3 + {\cal O}(R^{-4}),
\end{align}
where the subscripts indicate the corresponding powers of $R^{-1}$.
In the following,
to simplify the expansion
we use another coordinate transformation,
$s_1 = \rho \sin{\alpha}$ and 
$s_2 = \rho \cos{\alpha}$.
The scaled hyperangular volume element becomes
\begin{align}
\label{eq_scaledhyperangularvolumeelement}
d\bar{\bm{\Omega}} = &
R^{-2}  
\left\{ 1 - \frac{s_1^2 + s_2^2}{6 R^2} + {\cal O}(R^{-4})\right\}
ds_1 ds_2 
d\hat{\bm{\rho}}_1d\hat{\bm{\rho}}_2d\hat{\bm{\rho}}_3.
\end{align}
Integrals over the hyperangles in this asymptotic analysis 
are allowed to go to $\infty$ in $s_1$ and $s_2$, 
which introduces only exponentially small errors at large $R$ 
for the dimer-dimer channels under present consideration.

\subsection{Evaluating $H_0$}
The first order $H_0$ in the effective Hamiltonian 
is $R$-independent,
\begin{align}
\label{eq_H0}
H_0 & = h_1 + h_2,
\end{align}
where
\begin{align}
\label{eq_HCoulomb1}
h_j = -\frac{1}{2\mu}\frac{\partial^2}{\partial s_j^2} 
+ \frac{1}{2\mu } \frac{\bm{L}_j^2}{s_j^2} - \frac{Z_j}{s_j}
\end{align}
is the scaled hydrogenic Hamiltonian 
with ``charge'' $Z_j$;
\begin{align}
\label{eq_Z1}
Z_1=-q_1q_2\sqrt{\mu_{12}/\mu}
\end{align}
and 
\begin{align}
Z_2=-q_3q_4\sqrt{\mu_{34}/\mu}.
\end{align}
Note,
the first derivatives are removed due to scaling of the channel functions,
see e.g. Eq.~\eqref{eq_phiscaled} and the resulting
scaled hyperangular kinetic energy Eq.~\eqref{eq_lambda2}.
The first two terms of Eq.~\eqref{eq_HCoulomb1} come from
the expansion of Eq.~\eqref{eq_lambda2},
while the last term comes from the expansion of Eq.~\eqref{eq_C}.
The (bound) dimers form if the constituent dimer charges are of opposite sign.

Assuming the constituent dimer charges are point particles of opposite sign,
the unsymmetrized solutions to $H_0$ 
are a product of the 1-D scaled radial Coulomb solutions 
$u_{nl}$ and a set of coupled spherical harmonics.
For $s_j$, $j=1,2$, the radial solution $u_{nl}(x_j)$, 
normalized with respect to $ds_j$, 
is
\begin{align}
\label{eq_hydrogen}
u_{nl}(x_j) = \sqrt{\frac{\mu Z_j(n-l-1)!}{n^2 (n+l)!}}
\;e^{-\tfrac{x_j}{2}} x_j^{l+1} L_{n-l-1}^{2l+1}\!\!\left(x_j\right),
\end{align}
where
$x_j=2\mu Z_js_j/n$.
The quantum numbers range from $n=1,2,\ldots$ and $l=0,\ldots,n-1$.
Equation~\eqref{eq_hydrogen} solves
Eq.~\eqref{eq_HCoulomb1} with eigenvalues $-\mu Z_j^2/(2n^2)$,
such that
the zeroth-order energies $E_0$ are 
\begin{align}
E_0 = -\frac{\mu Z_1^2}{2 n_1^2} - \frac{\mu Z_2^2}{2 n_2^2}.
\end{align}

The full unsymmetrized solution $\bar{\Phi}_0(R;\bm{\Omega})$ at this order
is the product of radial solutions and spherical harmonics
coupled to a total angular momentum $L$ and projection $M$,
\begin{align}
\label{eq_phi0}
\bar{\Phi}_0 & (R;\bm{\Omega}) = {\cal N}(R) \;
u_{n_1l_1}(R \beta \sin \alpha) \times \nonumber \\
&
u_{n_2l_2}(R \beta \cos \alpha) 
{\cal Y}_{LM}^{\bm l}(\hat{\bm{\rho}}_1,\hat{\bm{\rho}}_2,\hat{\bm{\rho}}_3).
\end{align}
Here,
\begin{align}
\label{eq_YYY}
&{\cal Y}_{LM}^{\bm{l}}(\hat{\bm{\rho}}_1,\hat{\bm{\rho}}_2,\hat{\bm{\rho}}_3)
= \sum_{\bm{m}} 
\langle l_1m_1;l_2m_2|l_{12}m_{12}\rangle \times \nonumber \\
&\langle l_{12}m_{12};l_3m_3|LM \rangle
Y_{l_1m_1}(\hat{\bm{\rho}}_1)
Y_{l_2m_2}(\hat{\bm{\rho}}_2)
Y_{l_3m_3}(\hat{\bm{\rho}}_3),
\end{align}
where $\bm{l}$ and $\bm{m}$ represent the full set of $l$ and $m$
quantum numbers and
$\langle \cdot;\cdot|\cdot \rangle$ 
is a Clebsch-Gordan coefficient.
At this order, $R^{0}$,
the $\bm{l}$ and $\bm{m}$ are good quantum numbers,
but at higher orders in $R$, e.g. $R^{-2}$ or higher, they do not,
in general,
remain good quantum numbers.
The $R$-dependent normalization ${\cal N}(R)$ is straightforward to calculate
by demanding that
\begin{align}
&
\int d\bar{\bm{\Omega}} 
\bar{\Phi}_0^*(R;\bm{\Omega})\bar{\Phi}_0(R;\bm{\Omega})=1
\end{align}
whereby
\begin{align}
\label{eq_norm}
{\cal N} = R 
\bigg\{ &1 
  + \frac{  \langle n_1l_1|s_1^2|n_1l_1 \rangle 
          + \langle n_2l_2|s_2^2|n_2l_2 \rangle}{12 R^2}  
  + {\cal O}(R^{-4})
\bigg\}.
\end{align}
The terms $\langle |.| \rangle$ indicate standard
radial hydrogenic matrix elements~\cite{BetheSalpeter}.

Ignoring the $R$ prefactor of Eq.~\eqref{eq_norm} 
that cancels with that of the volume element 
Eq.~\eqref{eq_scaledhyperangularvolumeelement},
the volume element leads to effective off-diagonal elements
of $\langle H_0 \rangle$ of order $R^{-2}$
if evaluated in the basis $|\bar{\Phi}_0 \rangle$.
The normalization, on the other hand,
ensures that the diagonal elements of $\langle H_0 \rangle$ have corrections
of higher order, $R^{-4}$.
If this nonzero overlap due to the volume element is viewed as a perturbation,
then it contributes at second order in perturbation theory
leading to coupling at order $R^{-4}$.
These higher-order couplings are not treated in this work.

\subsection{Evaluating $H_1$}

The hyperradius $R$ first enters the asymptotic expansion
Eq.~\eqref{eq_expand2} 
of $H_{\rm ad}$ at first order,
\begin{align}
\label{eq_H1}
H_1 & = \frac{d_1}{R} 
\end{align}
with charge-mass coefficient $d_1$,
\begin{align}
\label{eq_d1}
d_1 = (q_1+q_2)(q_3+q_4)\sqrt{\mu_{dd}/\mu}.
\end{align} 
This order vanishes if either of the bound dimers are charge neutral,
e.g. if $q_2 \to -q_1$ or $q_4 \to -q_3$,
as this term represents coupling between ionic dimers
treated as point particles.
Nevertheless,
the zeroth order wave functions $\bar{\Phi}_0(R;\bm{\Omega})$ remain
the correct wave functions and,
treating $R$ as a parameter,
the action of $H_1$ is to shift all energies by the same amount.

The diagonal elements of $\langle H_1 \rangle$ are of order $R^{-2}$
when evaluated in the basis $|\bar{\Phi}_0 \rangle$.
Taking into account the volume element 
Eq.~\eqref{eq_scaledhyperangularvolumeelement}
and the normalization Eq.~\eqref{eq_norm},
the off-diagonal matrix elements 
of $\langle H_1 \rangle$ are of order $R^{-3}$
(and corrections to the diagonal matrix elements of order $R^{-5}$).
Similar to the effect in $\langle H_0 \rangle$,
if viewed as a perturbation,
these off-diagonal corrections contribute 
at second order in perturbation theory.
At that order,
there are couplings at order $R^{-6}$,
but these higher-order couplings are not treated in this work.

\subsection{Evaluating $H_2$}
At second order in Eq.~\eqref{eq_expand2},
\begin{align}
\label{eq_H2}
H_2 = H_2^{\bm \Lambda} + H_2^C,
\end{align}
where
\begin{align}
\label{eq_H2L}
& 
H_2^{\bm \Lambda} = \frac{1}{6 \mu R^2} 
\Bigg\{ 
    \bm{L}_1^2 \left(1+\frac{s_2^2}{s_1^{2}}\right) 
  + \bm{L}_2^2 \left(\frac{s_1^2}{s_2^{2}}+1\right) 
  + 3\bm{L}_3^2
\nonumber \\
 &- s_1^2\frac{\partial^2}{\partial s_2^2}
  - s_2^2\frac{\partial^2}{\partial s_1^2}
  + 2 s_1\frac{\partial}{\partial s_1}
  + 2 s_2\frac{\partial}{\partial s_2}
  + 2 s_1\frac{\partial}{\partial s_1}s_2\frac{\partial}{\partial s_2}
\Bigg\}
\end{align}
comes from the expansion of the hyperangular kinetic energy 
Eq.~\eqref{eq_lambda2}
and
\begin{align}
\label{eq_H2C}
H_2^C = &
\frac{1}{R^2} 
\Bigg\{ GD
  - \!\frac{Z_1\! }{6} \!\left( s_1 + \frac{s_2^2}{s_1} \right) 
  - \!\frac{Z_2\! }{6} \!\left( \frac{s_1^2}{s_2} + s_2\right) \!\!
\Bigg\}
\end{align}
comes from the expansion of the Coulomb potential Eq.~\eqref{eq_C}.
The terms contained within ``$GD$,''
\begin{align}
GD = 
  d_2^{(1)}  s_1 \cos \theta_{13}
+ d_2^{(2)}  s_2 \cos \theta_{23},
\end{align}
are the hyperradial analog of the first-order multipole expansion.
These terms lead to Gailitis-Damburg corrections~\cite{gailitis1963},
where the charge-mass coefficients $d_2^{(1)}$ and $d_2^{(2)}$ are
\begin{align}
\label{eq_d12}
d_2^{(1)}=(q_3+q_4)\left(\frac{q_2}{m_2}-\frac{q_1}{m_1}\right) 
  \mu_{dd} \sqrt{\mu_{12}/\mu}
\end{align}
and
\begin{align}
\label{eq_d34}
d_2^{(2)}= (q_1+q_2)\left(\frac{q_3}{m_3}-\frac{q_4}{m_4}\right) 
  \mu_{dd} \sqrt{\mu_{34}/\mu} \;.
\end{align}
As can be seen in Eqs.~\eqref{eq_d12} and~\eqref{eq_d34},
if a dimer is charge neutral,
then the corresponding first-order multipole expansion term vanishes. 
The charge to mass ratios of the dimer's constituent particles 
add together and scale the overall strength of these terms.
When the dissociating dimer has a nondegenerate energy eigenstates 
of definite parity, 
the GD terms vanish.  
But if opposite parity levels of a fragmenting dimer are degenerate, 
as is the case for all hydrogenic systems in excited states, 
the Gailitis-Damburg terms must be diagonalized 
using degenerate perturbation theory~\cite{greene1980}.

If evaluated in the basis $|\bar{\Phi}_0 \rangle$, 
$\langle H_2^{\bm \Lambda} \rangle$ remains diagonal
within a degenerate subspace. 
Said differently,
ignoring any accidental degeneracies
only states with the same $n_1$ and $n_2$
but different sets of $\bm{l}$ quantum numbers can couple.
However,
only the multipole expansion terms of $H_2^C$
introduce coupling within a degenerate manifold
and efficient methods exist to calculate such off-diagonal
matrix elements~\cite{Sanchez1992}.
Focusing on the diagonal elements $\langle \bar{\Phi}_0 | H_2 | \bar{\Phi}_0 \rangle$,
after standard hydrogenic integrals and much algebra yields
\begin{align}
\label{eq_H2diag}
&\langle \bar{\Phi}_0 | H_2 | \bar{\Phi}_0 \rangle =
\frac{l_3(l_3+1)}{2\mu R^2}
+\langle \bar{\Phi}_0 | GD | \bar{\Phi}_0 \rangle
\nonumber \\ &
+\frac{1}{4\mu R^2} 
\bigg\{\!\!-1-n_1^2+l_1(l_1+1)-n_2^2+l_2(l_2+1)
\bigg\}
\end{align}

Motivated by low energy scattering,
if we assume one of the dimers is in its ground state,
then the coupling angular integrations 
from the multipole expansion terms simplify.
In this case,
Eq.~\eqref{eq_YYY} reduces to either $|A_{23}\rangle$ or $|A_{13}\rangle$,
where
\begin{align}
\label{eq_A23}
|A_{23}\rangle =  \!\! \sum_{m_2m_3} \!\!\langle l_2m_2;l_3m_3 | LM \rangle
Y_{00}(\hat{\bm{\rho}}_1)
Y_{l_2m_2}(\hat{\bm{\rho}}_2)
Y_{l_3m_3}(\hat{\bm{\rho}}_3)
\end{align}
and
\begin{align}
\label{eq_A13}
|A_{13}\rangle = \!\!\sum_{m_1m_3} \!\!\langle l_1m_1;l_3m_3 | LM \rangle
Y_{l_1m_1}(\hat{\bm{\rho}}_1)
Y_{00}(\hat{\bm{\rho}}_2)
Y_{l_3m_3}(\hat{\bm{\rho}}_3).
\end{align}
At this order,
the only nonzero couplings are 
\begin{align}
\label{eq_angularcoupling}
\langle A_{k3}' | P_l(\cos \theta_{k3}) | A_{k3} \rangle 
= 
 c_k c_3 (-1)^L
 &\left\{ 
  \begin{array}{ccc}
  l_3' & l_k' & L \\
  l_k  & l_3  & l
  \end{array}
  \right\},
\end{align}
$k=1,2$,
with $l=1$ in the Legendre polynomial $P_l$.
Here,
\begin{align}
c_j = (-1)^{l_j} \sqrt{2l_j'+1} \langle l_j' 0; 1 0 | l_j 0 \rangle
\end{align}
and $\{\cdots\}$ is a Wigner six-J symbol.

\subsection{Evaluating $H_3$}
The third-order correction comes exclusively from 
expansion of the Coulomb potential, Eq.~\eqref{eq_C}.
\begin{align}
\label{eq_H3}
&H_3  = \frac{1}{R^3} 
\Bigg\{
  \frac{d_1}{2} \left(s_1^2+s_2^2\right) 
\\ &
- \frac{d_2^{(1)}d_2^{(2)}}{d_1} s_1 s_2 
    \big[ \cos \theta_{12} -3 \cos \theta_{13} \cos \theta_{23} \big]
\nonumber \\ &
- \frac{d_3^{(1)}}{2} s_1^2 \left[ 1 -3 \cos^2 \theta_{13} \right]
- \frac{d_3^{(2)}}{2} s_2^2 \left[ 1 -3 \cos^2 \theta_{23} \right]
\Bigg\} ,
\nonumber
\end{align}
where the charge-mass coefficients $d_1$, $d_2^{(1)}$, and $d_2^{(2)}$
are given in Eqs.~\eqref{eq_d1}, \eqref{eq_d12}, and~\eqref{eq_d34},
respectively,
while $d_3^{(1)}$ and $d_3^{(2)}$ are
\begin{align}
\label{eq_d31}
d_3^{(1)} = (q_3+q_4)\left(\frac{q_1}{m_1^2}+\frac{q_2}{m_2^2}\right) 
  \frac{\mu_{dd}^{3/2} \mu_{12}}{\mu^{1/2}} .
\end{align}
and 
\begin{align}
\label{eq_d32}
d_3^{(2)} = (q_1+q_2)\left(\frac{q_3}{m_3^2}+\frac{q_4}{m_4^2}\right) 
  \frac{\mu_{dd}^{3/2} \mu_{34}}{\mu^{1/2}} .
\end{align}
The first term of Eq.~\eqref{eq_H3} 
is unique to the hyperradial expansion.
It is not analogous to any term from the multipole
expansion in Jacobi coordinates.
It vanishes if either of the dimers are charge neutral,
or if non-vanishing,
leads to additional couplings between channels.
The last two terms are the second-order multipole expansion terms.
The multipole term in $\theta_{13}$, coming from the $(1,2)$ dimer, 
vanishes if the far $(3,4)$ dimer is charge neutral,
while 
the multipole term in $\theta_{23}$, coming from the $(3,4)$ dimer, 
vanishes if the far $(1,2)$ dimer is charge neutral.
The middle term of Eq.~\eqref{eq_H3} does not vanish
if the dimers are bound and represents two interacting dipoles.

For completeness,
assuming one dimer is in the ground state,
the angular coupling from the last two terms of Eq.~\eqref{eq_H3}
are given by Eq.~\eqref{eq_angularcoupling} with $l=2$.
The only nonzero angular coupling from the middle term is
\begin{align}
&\langle A_{13}'|\cos\theta_{12} 
               -3 \cos\theta_{13} \cos\theta_{23} |A_{23} \rangle =
\delta_{l_1'1}\delta_{l_21} 
\bigg\{ 
\frac{4\pi}{3} \delta_{l_3l_3'} - 
\nonumber \\
&
\frac{(-1)^{l_3+l_3'} }{2L+1} 
\sqrt{2l_3+1} \langle l_3 0 ; 1 0 |L 0 \rangle
\sqrt{2l_3'+1} \langle l_3' 0 ; 1 0 |L 0 \rangle
\bigg\}
\end{align}

\subsection{Lowest order nonadiabatic couplings}
To lowest order, 
the nonadiabatic couplings Eqs.~\eqref{eq_Pcoupling} and~\eqref{eq_Qcoupling}
can be calculated using the zeroth-order wave function Eq.~\eqref{eq_phi0},
neglecting the higher-order corrections to the normalization
and the volume element.
The derivatives with respect to the hyperradius $R$
do not affect the polar angles $\hat{\bm{\rho}}_j$ of the Jacobi vectors
and these integrations are trivial.
After performing the integration over the polar angles,
the $P$ matrix between dimer-dimer states is
\begin{align} 
P_{\nu \nu'} = &\delta_{\bm{l}\bm{l}'} \! \int R \;
u_{n_1l_1}(R \beta \sin \alpha) 
u_{n_2l_2}(R \beta \cos \alpha) 
\times \nonumber \\ &
\frac{\partial}{\partial R} R \;
u_{n_1'l_1}(R \beta \sin \alpha) 
u_{n_2'l_2}(R \beta \cos \alpha) \beta d\beta d\alpha
\nonumber \\ 
= &\frac{\delta_{\bm{l}\bm{l}'}}{R} 
\bigg\{
  \delta_{n_2n_2'}\langle n_1|s_1 \partial_1 |n_1' \rangle
 +\delta_{n_1n_1'}\langle n_2|s_2 \partial_2 |n_2' \rangle
\nonumber \\ &
 \qquad +\delta_{n_1n_1'}\delta_{n_2n_2'}
\bigg\}.
\end{align}
Here, $\partial_j$ represent derivatives with respect to $s_j$.
This long range coupling only occurs between 
dimer-dimer states of the same
internal angular momentum,
but different principle excitations.

The $Q$ matrix is 
\begin{align}
Q_{\nu\nu'}
= \frac{\delta_{\bm{l}\bm{l}'}}{R^2} 
\bigg\{ &
  2 \delta_{n_2n_2'}\langle n_1|s_1 \partial_1 |n_1' \rangle
 +2 \delta_{n_1n_1'}\langle n_2|s_2 \partial_2 |n_2' \rangle
\nonumber \\ &
 +  \delta_{n_2n_2'}\langle n_1|s_1^2 \partial_1^2 |n_1' \rangle
 +  \delta_{n_1n_1'}\langle n_2|s_2^2 \partial_2^2 |n_2' \rangle
\nonumber \\ &
 +2 \langle n_1|s_1 \partial_1 |n_1' \rangle 
    \langle n_2|s_2 \partial_2 |n_2' \rangle 
\bigg\}.
\end{align}
The diagonal elements are
\begin{align}
\frac{Q_{\nu\nu}}{2\mu}
= \frac{1}{4\mu R^2} 
\bigg\{\!\!-1-n_1^2+l_1(l_1+1)-n_2^2+l_2(l_2+1)
\bigg\},
\end{align}
which can be seen match with the terms in curly brackets
from Eq.~\eqref{eq_H2diag}.
Thus,
taking the adiabatic potentials with nonadiabatic diagonal correction,
the only remaining terms are
an angular momentum barrier due to the inter-dimer angular momentum
$l_3$ and the first-order multipole corrections.
One key result of the present analysis 
is analogous to what Macek found in the three-body system~\cite{macek1968}, 
namely that the combination of $U_{\nu}(R)-\frac{1}{2\mu}Q_{\nu,\nu}(R)$ 
at long range has the expected coefficient of $1/2\mu R^2$, 
namely $l_3(l_3+1)$ 
(or the eigenvalues of the Gailitis-Damburg corrections 
in cases where they contribute 
in first-order degenerate perturbation theory).  
This is reassuring because it means that the asymptotic centrifugal barrier 
has the same coefficient in hyperspherical coordinates 
as it has in conventional Jacobi coordinates.

\subsection{Special cases}
An important case of fragmenting bound dimers
is the case where they are identical.
That is,
if the equally-charged particles are identical fermions,
then $m_1=m_3=m_A$ and $q_1=q_3=q_A$, and $m_2=m_4=m_B$ and $q_2=q_4=q_B$.
In this case the wave function must be antisymmetric with respect to
the two ``A'' particles and the two ``B'' particles.
The corresponding antisymmetrizing operator is $1-P_{13}-P_{24}+P_{13}P_{24}$
acting on the basis $|\bar{\Phi}_0\rangle$.
The results $P_{13}|\bar{\Phi}_0\rangle$ and $P_{24}|\bar{\Phi}_0\rangle$,
however,
are exponentially suppressed due to the spatial part of the wave function
in the asymptotic limit.
This can be seen since these permutations rotate $\bm{\rho}_1$
and $\bm{\rho}_2$ such that they pick up components along all Jacobi vectors.
Thus, in the asymptotic limit,
the arguments of the hydrogenic wave functions are proportional to the
asymptotically large hyperradius.
To a good approximation then the antisymmetrization is accomplished
by $(1+P_{13}P_{24})|\bar{\Phi}_0\rangle$,
where the action of the last operator 
is to swap $\bm{\rho}_1$ and $\bm{\rho}_2$ 
and reverse the direction of $\bm{\rho}_3$.

Because the dimers are identical,
there are also additional degeneracies to consider.
For example,
in general 
$\langle \bar{\Phi}_0| H_2 |P_{13}P_{24} \bar{\Phi}_0\rangle \ne 0$ 
and degenerate perturbation theory is required.
This is true even for charge-neutral dimers,
though the multipole expansion terms and any coupling thereof 
would be eliminated.
For charge-neutral identical dimers,
there is an additional symmetry where the system is invariant
under charge conjugation.
The charge conjugation projection operator $\hat{C}$,
which can be written here as 
$\hat{C}=(1+P_{12}P_{34})$,
reverses the direction of $\bm{\rho}_1$ and $\bm{\rho}_2$,
but leaves $\bm{\rho}_3$ unchanged.
Thus,
the antisymmetrized channel functions $(1+P_{13}P_{24})|\bar{\Phi}_0\rangle$ 
are already eigenstates
of $\hat{C}$ with eigenvalues $(-1)^{l_1+l_2}$.

The equal mass and charge case includes the positronium dimer system,
which is studied in more detail in Ref.~\cite{daily2014}.

\section{Particle-trimer configuration}
\label{sec_pt}
\subsection{Coordinate system and Coulomb interaction}
To describe the break up into a trimer and free particle,
it is convenient to choose the K-type set of Jacobi vectors.
The relative coordinates are defined via the coordinate transformation,
conveniently written in matrix form, as
\begin{align}
\label{eq_jmatrix_pt}
\begin{pmatrix} 
\bm{\rho}_1 \\ \bm{\rho}_2 \\ \bm{\rho}_3 \\ \bm{R}_{CM}
\end{pmatrix} \!= \!
\left( \!\!\!\!
\begin{array}{rlccr}
\sqrt{\frac{\mu_{12}}{\mu}}  \times & \{ 1 & -1 & 0 & 0 \} \\
\sqrt{\frac{\mu_{123}}{\mu}} \times &\{\frac{\mu_{12}}{m_2} 
                                   & \frac{\mu_{12}}{m_1} 
                                   & -1 & 0 \} \\
\sqrt{\frac{{\mu_{\rm pt}}}{\mu}}\times &\{\frac{m_1}{M_3} 
                                   &\frac{m_2}{M_3} 
                                   &\frac{m_3}{M_3} 
                                   & -1 \} \\
\frac{1}{M_4} \times &\{ m_1 & m_2 & m_3 & m_4 \}
\end{array} \right) \!\!\!\!
\begin{pmatrix}
\bm{r}_1 \\ \bm{r}_2 \\ \bm{r}_3 \\ \bm{r}_4
\end{pmatrix} \!\! ,
\end{align}
where first $\bm{r}_1$ is joined to $\bm{r}_2$ 
(which defines $\bm{\rho}_1$),
then the center of mass of this pair is joined to $\bm{r}_3$
(which defines $\bm{\rho}_2$),
and last the center of mass of the trimer cluster  
is joined to $\bm{r}_4$ (which defines $\bm{\rho}_3$).
$M_3$ and $M_4$ are defined in Eq.~\eqref{eq_massn} and
the newly introduced reduced masses are 
the atom-dimer mass $\mu_{\rm pt}$,
${\mu_{\rm pt}}=M_3 m_4/M_4$, and
$\mu_{123}=M_2 m_3/M_3$.
Note,
though all equations in Sec.~\ref{sec_theory} still apply,
the hyperangles of this configuration are different
from the dimer-dimer configuration.

The Coulomb part $V_C$ is straigtforward to calculate in Jacobi coordinates,
where the matrix from Eq.~\eqref{eq_jmatrix_pt} is first inverted to
define the $\bm{r}_j$ in terms of the $\bm{\rho}_j$.
\begin{widetext}
\begin{align}
\label{eq_Vc_pt}
V_C &= 
   q_1q_2 \frac{\sqrt{\mu_{12} / \mu}}{\rho_1}
  +q_1q_3 \left|  \frac{\sqrt{\mu\;\mu_{12}}}{m_1}\bm{\rho}_1 
                + \sqrt{\frac{\mu}{\mu_{123}}}\bm{\rho}_2 \right|^{-1}
  +q_1q_4 \left|  \frac{\sqrt{\mu\;\mu_{12}}}{m_1}\bm{\rho}_1 
                + \frac{\sqrt{\mu\;\mu_{123}}}{M_2}\bm{\rho}_2 
                + \sqrt{\frac{\mu}{\mu_{\rm pt}}}\bm{\rho}_3 \right|^{-1} 
\nonumber \\
& +q_2q_3 \left|  \frac{\sqrt{\mu\;\mu_{12}}}{m_2}\bm{\rho}_1 
                - \sqrt{\frac{\mu}{\mu_{123}}}\bm{\rho}_2 \right|^{-1}
  +q_2q_4 \left|  \frac{\sqrt{\mu\;\mu_{12}}}{m_2}\bm{\rho}_1 
                - \frac{\sqrt{\mu\;\mu_{123}}}{M_2}\bm{\rho}_2 
                - \sqrt{\frac{\mu}{\mu_{\rm pt}}}\bm{\rho}_3 \right|^{-1}
  +q_3q_4 \left|  \frac{\sqrt{\mu\;\mu_{123}}}{m_3} \bm{\rho}_2
                - \sqrt{\frac{\mu}{\mu_{\rm pt}}} \bm{\rho}_3 \right|^{-1}.
\end{align}
Expressing Eq.~\eqref{eq_Vc_pt} in hyperspherical coordinates yields
\begin{align}
\label{eq_C_pt}
C(\bm{\Omega}) = 
& q_1q_2 \frac{\sqrt{\mu_{12}/\mu}}{\sin \alpha\sin \beta} + \nonumber \\
&+q_1q_3 \bigg[    \frac{\mu\;\mu_{12}}{m_1^2}\sin^2 \alpha\sin^2 \beta 
                 + \frac{\mu}{\mu_{123}}   \cos^2 \alpha\sin^2 \beta 
                 + \frac{\mu\;\sqrt{\mu_{12}}}{m_1 \sqrt{\mu_{123}}} 
                        \sin(2\alpha)\sin^2 \beta\cos \theta_{12} \bigg]^{-1/2}
\nonumber \\
&+q_1q_4 \bigg[    \frac{\mu\;\mu_{12}}{m_1^2}\sin^2 \alpha\sin^2 \beta 
                 + \frac{\mu\;\mu_{123}}{M_2^2}\cos^2 \alpha\sin^2 \beta 
                 + \frac{\mu}{{\mu_{\rm pt}}}\cos^2 \beta
                 + \frac{\mu\;\sqrt{\mu_{12}\mu_{123}}}{m_1M_2} 
                        \sin(2\alpha)\sin^2 \beta\cos \theta_{12} \nonumber \\
& \qquad\qquad   + \frac{\sqrt{\mu_{12}}\;\mu}{\sqrt{{\mu_{123}}}\;m_1}
                        \sin \alpha\sin(2\beta)\cos \theta_{13}
                 + \frac{\sqrt{\mu_{123}}\;\mu}{\sqrt{{\mu_{\rm pt}}}\;M_2}
                        \cos \alpha\sin(2\beta)\cos \theta_{23}
         \bigg]^{-1/2}
\nonumber \\
&+q_2q_3 \bigg[    \frac{\mu\;\mu_{12}}{m_2^2}\sin^2 \alpha\sin^2 \beta 
                 + \frac{\mu}{\mu_{123}}   \cos^2 \alpha\sin^2 \beta 
                 - \frac{\mu\;\sqrt{\mu_{12}}}{m_2 \sqrt{\mu_{123}}} 
                        \sin(2\alpha)\sin^2 \beta\cos \theta_{12} \bigg]^{-1/2}
\nonumber \\
&+q_2q_4 \bigg[    \frac{\mu\;\mu_{12}}{m_2^2}\sin^2 \alpha\sin^2 \beta
                 + \frac{\mu\;\mu_{123}}{M_2^2}\cos^2 \alpha\sin^2 \beta 
                 + \frac{\mu}{{\mu_{\rm pt}}}\cos^2 \beta
                 - \frac{\mu\;\sqrt{\mu_{12}\mu_{123}}}{m_1M_2} 
                        \sin(2\alpha)\sin^2 \beta\cos \theta_{12} \nonumber \\
& \qquad\qquad   - \frac{\sqrt{\mu_{12}}\;\mu}{\sqrt{{\mu_{\rm pt}}}\;m_1}
                        \sin \alpha\sin(2\beta)\cos \theta_{13}
                 + \frac{\sqrt{\mu_{123}}\;\mu}{\sqrt{{\mu_{\rm pt}}}\;M_2}
                        \cos \alpha\sin(2\beta)\cos \theta_{23}
         \bigg]^{-1/2}
\nonumber \\
&+q_3q_4 \bigg[    \frac{\mu\;\mu_{123}}{m_3^2} {\cos^2 \alpha\sin^2 \beta}
                 + \frac{\mu}{\mu_{\rm pt}} \cos^2 \beta 
                 - \frac{\sqrt{\mu_{123}}\;\mu}{\sqrt{\mu_{\rm pt}}\; m_3}
                        \cos \alpha \sin(2\beta) \cos \theta_{23} \bigg]^{-1/2}.
\end{align}
\end{widetext}

\subsection{Effective Hamiltonian}
The simultaneous limits $R\to\infty$ and $\beta \to 0$,
such that $R\beta = \rho$,
describe the situation where the centers of mass of 
the (1,2,3) trimer and the remaining particle 4 are far apart.
The asymptotic expansion of $H_{\rm ad}$ around 
$R \to \infty$, where the terms are grouped in powers of $1/R$, 
defines the effective Hamiltonian $H_{\rm eff}$,
\begin{align}
\label{eq_expand_pt}
H_{\rm eff} \approx H_0 + H_1 + H_2 + {\cal O}(R^{-3}).
\end{align}
As was done in Sec.~\ref{subsec_dd_eh},
to simplify the expansion
we use another coordinate transformation,
$s_1 = \rho \sin{\alpha}$ and 
$s_2 = \rho \cos{\alpha}$
resulting in the a scaled hyperangular volume element
of the same form as Eq.~\eqref{eq_scaledhyperangularvolumeelement}.

The first order $H_0$ in the effective Hamiltonian Eq.~\eqref{eq_expand_pt}
is $R$-independent, 
where
\begin{align}
\label{eq_H0_pt}
H_0 & =
-\frac{1}{2\mu}\frac{\partial^2}{\partial s_1^2} 
+ \frac{1}{2\mu}\frac{\bm{L}_1^2}{s_1^2}
-\frac{1}{2\mu}\frac{\partial^2}{\partial s_2^2} 
+ \frac{1}{2\mu}\frac{\bm{L}_2^2}{s_2^2}
+\sum_{i<j}^3 V_{ij}.
\end{align}
The interaction terms are
\begin{align}
V_{12} = &  q_1q_2\frac{\sqrt{\mu_{12}/\mu}}
                      {s_1}, 
\\
V_{13} = & q_1q_3\frac{m_1 \sqrt{\mu_{123}/\mu}}
                     {|\mu_3 \bm{s}_1+m_1\bm{s}_2|},
\end{align}
and
\begin{align}
V_{23} = q_2q_3 \frac{m_2 \sqrt{\mu_{123}/\mu}}
                    {|\mu_3 \bm{s}_1-m_2\bm{s}_2|},
\end{align}
where
\begin{align}
  \mu_3 = \left( \frac{m_1m_2m_3}{m_1+m_2+m_3} \right)^{1/2}.
\end{align}
Equation~\eqref{eq_H0_pt} is recognizable 
as the scaled relative Coulomb three-body Hamiltonian,
which has no known analytical solution 
but has been previously solved numerically to high accuracy using a number
of approaches (see e.g. Refs.~\cite{2013RMPmitroy,ancarani2011,
abramov1996,flores1998} and references therein).
The zeroth-order energies can be found in the suggested references
and elsewhere.

The hyperradius $R$ first enters the asymptotic expansion
Eq.~\eqref{eq_expand_pt} at first order,
\begin{align}
H_1 = \frac{1}{R} (q_1+q_2+q_3) q_4 \sqrt{\mu_{\rm pt}/\mu}.
\end{align}
This order vanishes if the trimer is charge neutral,
as this correction represents the ionic potential
between the trimer and free particle.
Similar to the dimer-dimer case,
the effect of $H_1$ is to shift all energies by the same amount.

At second order,
\begin{align}
H_2 = H_2^{\bm{\Lambda}} + H_2^C
\end{align}
where $H_2^{\bm{\Lambda}}$ is of the same form as
Eq.~\eqref{eq_H2L} and comes from the expansion of the 
hyperangular kinetic energy Eq.~\eqref{eq_lambda2}.
$H_2^C$,
\begin{align}
\label{eq_H2C_pt}
H_2^C = &\frac{1}{6 R^2} \left(s_1^2+s_2^2\right)
                          \left(V_{12}+V_{13}+V_{23}\right)
\nonumber \\ &
+ d_2^A s_1 \cos \theta_{13}
+ d_2^B s_2 \cos \theta_{23},
\end{align}
comes from expansion of the Coulomb terms of Eq.~\eqref{eq_C_pt}.
The charge-mass coefficients $d_2^A$ and $d_2^B$ are
\begin{align}
d_2^A = q_4 \left(\frac{q_2}{\sqrt{\mu_{\rm pt}}} 
               - \frac{q_1}{\sqrt{\mu_{123}}}\right)
\frac{\mu_{12}^{1/2}\mu_{\rm pt}^{3/2}}{m_1\mu^{1/2}}
\end{align}
and
\begin{align}
d_2^B = q_4 \left(\frac{q_3}{m_3} - \frac{q_1+q_2}{M_2}\right)
\frac{\mu_{123}^{1/2} \mu_{\rm pt}}{\mu^{1/2}}.
\end{align}
The last two terms of Eq.~\eqref{eq_H2C_pt}
are analogous to the first order terms of a multipole expansion
of a charge distribution (the trimer) with a free charge.
Our expectation is that the sum of the large-$R$ adiabatic potential curve
$U_{\nu}(R)$ and the $\frac{-1}{2\mu} Q_{\nu\nu}(R)$ terms
cancel the first terms of Eq.~\eqref{eq_H2C_pt} and all but the angular 
momentum barrier in $l_3$ of $H_2^{\bm{\Lambda}}$.

\section{Conclusion and Outlook}
\label{sec_conclusions}
This paper derives the asymptotic Hamiltonian in powers of $R^{-1}$
for four-body systems of charges for both the dimer-dimer
and particle-trimer dissociation configurations.
The terms in the effective Hamiltonian are identified with
analogous terms in the multipole expansion,
with some complications associated with nonadiabatic couplings 
in the hyperspherical representation.
For the dimer-dimer configuration,
diagonal matrix elements through order $R^{-2}$ were calculated
and off-diagonal couplings were discussed at this same order and higher.
For the particle-trimer configuration,
only up through second-order corrections in the effective Hamiltonian
are derived.

The charged four-body system with arbitrary masses 
constitutes a large parameter space.
In the adiabatic hyperspherical framework,
this work lends to the understanding of the dissociation of many systems
such as H$_2$, Ps$_2$, three-electron atoms or other exotic muonic 
four-body clusters. 
Having analytical knowledge of the long-range tail of the potential
curves also provides a check on numerical calculations
of the adiabatic potentials and their nonadiabatic corrections.

\section{Acknowledgements}
Support by the National Science Foundation through Grant No. PHY-1306905
and by the US Deptartment of Energy, Office of Science through Grant No. 
DE-SC0010545 is gratefully acknowledged.
We also acknowledge fruitful discussions with Chris. H. Greene.

\end{document}